\documentclass{elsart}
\usepackage{graphicx}
\usepackage{boxedminipage}\usepackage[dvips]{color}
\newcommand{\Vect}[1]{{\textnormal{\mathversion{bold}$#1$}}}
\newcommand{\dei}{{\rm d}}

\newcommand{\im}{{\rm i}}

\begin{document}
\begin{frontmatter}
 \title{Absence of the Effects of Vortices in the Gauge Glass}
 \author{Toshiyuki Hamasaki and Hidetoshi Nishimori}
 \address{Department of Physics, Tokyo Institute of Technology,
 Oh-okayama, Meguro-ku, Tokyo 152-8551, Japan}
 \begin{abstract}
  We calculate several correlation functions and distribution functions
  of dynamical variables for the gauge glass and the Villain model using
  the spin wave approximation and the gauge transformation.
  The results show that the spin wave approximation gives the exact
  solutions on the Nishimori line in the phase diagram.
  This implies that vortices play no role in the thermodynamic behavior
  of the system as long as some correlation and distribution functions
  are concerned.
  These results apply to any dimensions including the two-dimensional
  case.
 \end{abstract}
\end{frontmatter}

 \section{Introduction}
 The gauge glass, which is an extension of the $XY$ model to the random
 spin system, attracts both theoretical and practical interest.
 This model provides good descriptions of some physical situations, for
 example, the $XY$ magnet with random Dzyaloshinskii-Moriya interactions
 \cite{RSN83} and a granular superconductivity and Josephson-junction
 arrays with positional disorder \cite{GK86}\cite{SCS84}.
 Especially, the two-dimensional gauge glass has been studied actively
 because of the interesting relation between the Kosterlitz-Thouless
 (KT) transition and the effect of disorder.
 This relation was first discussed by Rubinstein, Shraiman and Nelson
 \cite{RSN83}.
 They have shown that for small amount of randomness, as the temperature
 is decreased, there appears first the paramagnetic phase, then a KT-like
 phase, and finally again a paramagnetic phase.
 Thus a reentrant transition appears in the gauge glass.
 After their work, Natterman, Scheidl, Korshunov and Li \cite{NSKL95}
 corrected their description of the reentrant transition to the one
 without reentrance.
 These latter authors pointed out an overestimation of vortex pair
 density in the previous work.
 Furthermore, Sheidl \cite{S97} found a new ordered phase in which 
 single vortex excitations occur in the reentrant phase, and Maucourt and
 Grempel \cite{MG97} suggested from Monte Carlo simulations that there
 is no indication of a low-temperature reentrant phase.\\
 Today, the two-dimensional gauge glass at low temperatures is generally
 believed to have no reentrance \cite{CD00}, except for some researchers
 \cite{MW99}.
 Hence, the present major interest in the two-dimensional gauge glass is
 the structure of the KT phase, namely, the boundary between the
 freezing phase of the vortex-pair excitations and the non-freezing
 phase.\\
 The gauge glass has also been investigated in the field of the spin
 glass theory because the gauge glass is an extension of the Ising spin
 glass with continuous spin variables \cite{N81}.
 Particularly, the method of gauge transformation, which was
 developed in the study of the spin glass, is a powerful technique for
 deriving analytical results for gauge glass.
 Ozeki and Nishimori \cite{ON93} found the exact solution of the
 internal energy of the gauge glass under a special condition using
 the method of gauge transformation.
 This special condition relevant to the exact solution corresponds to a
 line in the phase diagram called the Nishimori line. 
 Ozeki and Nishimori also showed that the phase boundary between the
 ordered and the disordered phases runs parallel to the temperature
 axis. 
 Although they could obtain the exact solution on the Nishimori line,
 there exists a mysterious property associated with the line:
 For example, the exact solution of the internal energy on the Nishimori
 line has no singularity as a function of temperature although the
 the line runs across the phase boundary.
 From these facts in mind, we aim to clarify what occurs under
 the special condition of this line. 

 In this paper, we calculate some gauge invariant quantities on the
 Nishimori line.
 First, we define the gauge glass and show that the exact solution
 of the internal energy can be calculated under the special condition in
 Section 2.
 Next, we introduce the gauge invariant correlation functions and 
 calculate them using the spin wave approximation, gauge transformation
 and the Villain model in Section 3.
 The gauge invariant distribution functions are introduced and
 calculated in Section 4.
 From the results of Sections 3 and 4, we discuss the physical
 properties of the Nishimori line in Section 5.

 \section{General properties}
  \subsection{Gauge glass}
  The gauge glass is defined by the Hamiltonian
  \begin{eqnarray}
   H = -J\sum_{\left\langle i,j \right\rangle} 
    \cos (\theta_i -\theta_j -A_{ij}),
    \label{eqn:HGG}
  \end{eqnarray}
  where $A_{ij}$ is the quenched random phase shift and the coupling
  constant $J$ is positive.
  The sum runs over all nearest-neighbor pairs on a lattice.
  Note that we do not specify the spatial dimensionality explicitly
  here. 
  We assume that $A_{ij}$ is independently distributed at each
  bond as follows,
  \begin{eqnarray}
   P(A_{ij}) = \frac{1}{2\pi I_0(K_p)}{\rm e}^{K_p\cos A_{ij}},
    \label{dist:HGG}
  \end{eqnarray}
  where the magnitude of $K_p$ controls the tendency toward
  ferromagnetism;
  the system becomes the pure ferromagnetic $XY$ model in the limit of
  large $K_p$.
  The function $I_0(x)$ appearing in the denominator of eqn
  (\ref{dist:HGG}) is the modified Bessel function
  $I_0(x) = \int_{0}^{2\pi}\dei\theta\,\e^{x\cos\theta}$.
  Note that we express the thermal average by angular brackets
  $\langle\cdots\rangle$ and the average over the distribution of
  quenched randomness by square brackets $[\cdots]$.
  
  \subsection{Exact solution of the internal energy}
  The gauge transformation is a powerful method to calculate the exact
  solution of the internal energy for the gauge glass \cite{N01}.
  We introduce the gauge transformation to our model and derive the
  exact internal energy under a special condition.\\
  The gauge transformation is defined as 
  \begin{eqnarray}
   \theta_i \rightarrow \theta_i -\phi_i, \hspace*{5mm}
    A_{ij} \rightarrow A_{ij} -\phi_i +\phi_j,\label{eqn:gauge-TR}
  \end{eqnarray}
  where $\phi_i$ is a gauge variable fixed arbitrarily at each site.
  The Hamiltonian (\ref{eqn:HGG}) is invariant under this transformation.
  The internal energy is written explicitly as
  \begin{eqnarray}
   E = \big[\left\langle H \right\rangle \big] 
   =
    \int
    \prod_{\left\langle i,j \right\rangle}
    \frac{{\rm d}A_{ij}}{2\pi}
    \frac{\e^{K_p\cos A_{ij}}}{I_0(K_p)}\cdot 
    \frac
    {{\displaystyle\int\prod_i
    \frac{{\rm d}\theta_i}{2\pi}}
    H {\rm e}^{K\sum_{\left\langle i,j \right\rangle}
    \cos(\theta_i -\theta_j -A_{ij})}}
    {{\displaystyle\int\prod_i
    \frac{{\rm d}\theta_i}{2\pi}}
    {\rm e}^{K\sum_{\left\langle i,j \right\rangle}
    \cos(\theta_i -\theta_j -A_{ij})}},\label{eqn:first_eqn}
  \end{eqnarray}
  where the ranges of integration over $A_{ij}$ and $\theta_i$ are
  from $0$ to $2\pi$.
  Here we apply the gauge transformation to eqn (\ref{eqn:first_eqn}).
  Since the gauge transformation is just a change of running variables,
  the value of eqn (\ref{eqn:first_eqn}) is independent of the choice of
  $\{\phi_i\}$ and therefore we may integrate eqn (\ref{eqn:first_eqn})
  over all gauge variables $\{\phi_i\}$, each from 0 to $2\pi$, and
  divide each integration of $\phi_i$ by  $2\pi$.
  Using the special condition $K=K_p$, we obtain the exact internal
  energy as 
  \begin{eqnarray}
   E =  -JN_B\frac{I_1(K_p)}{I_0(K_p)}.\label{eqn:exact1}
  \end{eqnarray}
  Here the total number of interacting pairs is denoted as $N_B$.
  The function $I_1(x)$ is the modified Bessel function 
  $I_1(x) = \int_{0}^{2\pi}\dei\theta\cos\theta\,
   \e^{x\cos\theta}.$\\
  We could carry out the calculation of the internal energy
  exactly owing to the gauge invariance of the Hamiltonian and the
  special condition $K = K_p$.
  The gauge invariance plays an important role in the gauge glass or
  random spin system because, for the gauge invariant quantities,
  the exact solution similar to the case of the internal energy can be
  calculated by using gauge transformation.
  \begin{figure}[t]
   \begin{center}
    \includegraphics[width=7cm]{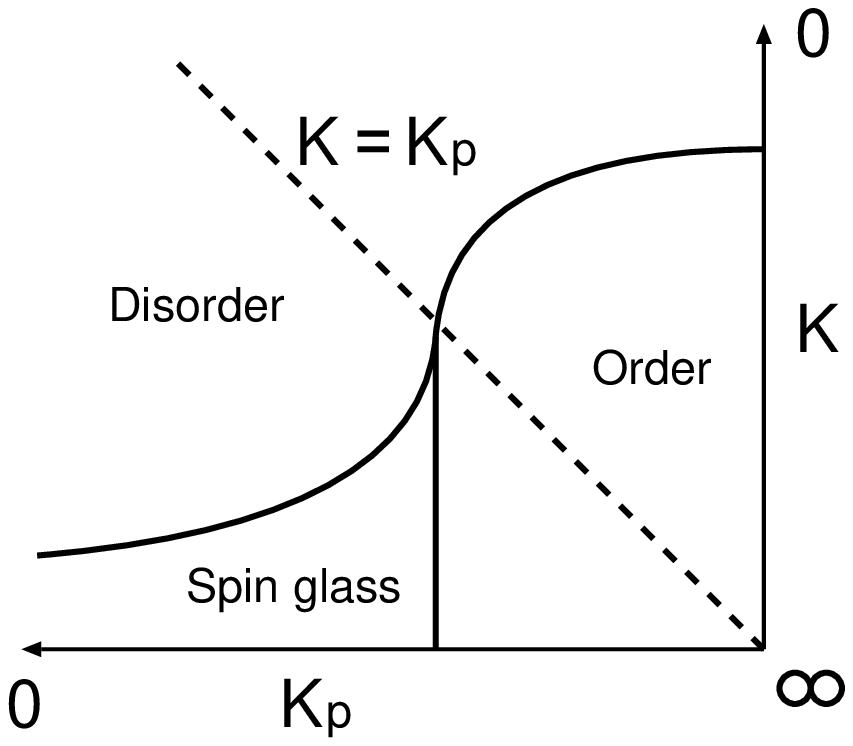}
    \caption{The schematic phase diagram and the Nishimori line (dashed).
    The spin glass phase exists in three dimensions or over.}
    \label{fig:nishiline}
   \end{center}
  \end{figure}
  The condition $K=K_p$ defines a line (the Nishimori line) in the phase
  diagram as in Figure \ref{fig:nishiline}.
  Although the line $K=K_p$ clearly runs across the phase boundary, the
  exact solution of the internal energy (\ref{eqn:exact1}) is a simple
  function because the modified Bessel functions has no singularity.
  Thus, we may consider that the singular part of the internal energy
  vanishes on the Nishimori line.
  The same behavior of the internal energy as in the gauge glass appears
  in the Gaussian spin glass model and  $\pm J$ spin glass model
  \cite{N01}.\\
  We can further evaluate some quantities using the spin wave
  approximation, which is a characteristic method in continuous spin
  systems, and obtain some information not available in Ising spin
  systems, which helps us to understand the significance of the
  Nishimori line.
  In particular, it is believed that the vortices created by the
  periodicity of eqn (\ref{eqn:HGG}) cause the singularity relevant to
  the phase transition in the two-dimensional gauge glass.
  Accordingly, we expect that a new physical picture relevant to
  the Nishimori line may be found by using the characteristic method in
  the continuous spin system.
  This program is carried out in the next section.

 \section{Gauge invariant correlation functions}
  \subsection{Definitions}
  In order to investigate the physical significance of the result in the
  previous section, we introduce the gauge invariant correlation
  functions as
  \begin{eqnarray}
   C_{\rm p} &=& \left[\left\langle 
			   \exp{\im(\theta_0 -\theta_r 
			   -\sum_{k=1}^{r} A_{k-1,k})} \right\rangle
		   \right]
   \label{eqn:c2}\\
   C_{\rm le} &=& \bigg[\Big\langle 
    \cos(\theta_0 -\theta_1 -A_{01})
    \cos(\theta_{r} -\theta_{r+1} -A_{r,r+1}) 
    \Big\rangle\bigg]
    \label{eqn:c3}\\
   C_{\rm ch} &=& \bigg[\Big\langle 
    \sin(\theta_0 -\theta_1 -A_{01})
    \sin(\theta_{r} -\theta_{r+1} -A_{r,r+1}) 
    \Big\rangle\bigg].
    \label{eqn:c4}
  \end{eqnarray}
  
\begin{figure}[t]
   \begin{center}
    \includegraphics*[width=15cm]{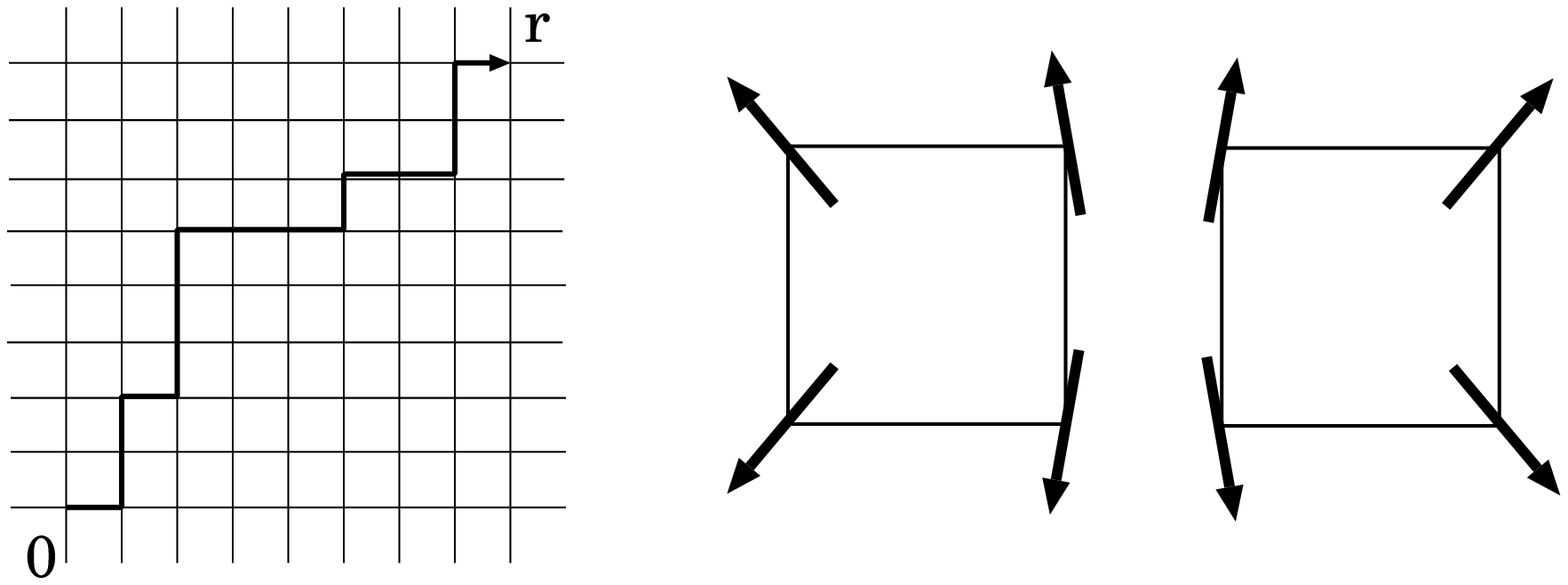}
    \caption{Schematic diagram of the correlation function $C_{\rm p}(r)$ and
    chirality. Left figure shows a path from site $0$ to site
    $r$. The right figure shows two states with different
    chiralities.}\label{fig:corre_fig}
   \end{center}
  \end{figure}

  Equation (\ref{eqn:c2}) is the gauge-invariant correlation function  
  which measures the correlation along a path from site $0$ to site $r$
  taking the phase twist of $A_{ij}$ into consideration. 
  In Figure \ref{fig:corre_fig}, we sketch a schematic picture of a path
  of correlation function $C_{\rm p}$.
  Equation (\ref{eqn:c3}) means the local energy correlation function.
  The local energy correlation function is also gauge invariant.
  Equation (\ref{eqn:c4}) represents the chirality correlation function.
  Chirality introduced by Villain \cite{V77} is an Ising-like degree of
  freedom.
  The chirality distinguishes the two states which are transformed to
  each other using the mirror transformation (Figure \ref{fig:corre_fig}).
  The chirality correlation function is also gauge invariant.

  \subsection{Spin wave approximation}
  Let us now consider the Hamiltonian introduced in eqn (\ref{eqn:HGG})
  in two dimensions.
  At sufficiently low temperatures, the effect of thermal fluctuations
  is small, so that the spins tend to align to each other with the
  twist $A_{ij}$ taken into account.
  Hence, we are able to expand the cosine term near $(\theta_i -\theta_j
  -A_{ij}) = 0$,
  \begin{eqnarray}
   H \simeq \frac{J}{2}
    \sum_{\langle i,j \rangle}(\theta_i -\theta_j -A_{ij})^2 
    \label{eqn;sw}
  \end{eqnarray}
  We note that this approximation neglects periodicity of cosine in eqn
  (\ref{eqn:HGG}).
  It is generally believed that, if we want to describe the phase
  transition in two-dimensional gauge glass, we must consider the
  contribution of periodicity or vortices
  \cite{CD00}.
  To ensure the validity of the expansion of eqn (\ref{eqn;sw}), we must
  assume that the phase twist $A_{ij}$ is sufficiently small, so that eqn
  (\ref{dist:HGG}) is expressed as
  \begin{eqnarray}
   P(A_{ij}) \propto {\rm e}^{-K_p A_{ij}^2/2}.
    \label{eqn:bond:gauss}
  \end{eqnarray}
  The constant $K_p^{-1}$ is equal to the variance of the Gaussian
  distribution $[A_{ij}^2]\equiv\sigma$.
  In the same way as in the spin wave approximation, the distribution of
  eqn (\ref{eqn:bond:gauss}) neglects periodicity of $A_{ij}$.
  This approximation of the distribution of $A_{ij}$ is expected to be
  sufficient to describe correlations at low temperature.
  
  Using the spin wave approximation of eqns (\ref{eqn;sw}) and
 (\ref{eqn:bond:gauss}), we can calculate the gauge invariant
 correlation functions as 
  \begin{eqnarray}
   C_{\rm p} &=& 
    \exp\left\{\frac{1}{2}
	 \left(\frac{1}{K_p} -\frac{1}{K}\right)
	 G_{0r} -\frac{\sigma r}{2}\right\}\\
   C_{\rm le} &=& 
    \frac{1}{2}
    \exp\left\{
	 \left(\frac{1}{K_p} -\frac{1}{K}\right)
	 G_{0r}^{(-)} 
	 -\sigma\right\}
    + 
    \frac{1}{2}
    \exp\left\{
	 \left(\frac{1}{K_p} -\frac{1}{K}\right)
	 G_{0r}^{(+)} 
	 -\sigma\right\}\\
   C_{\rm ch} &=& 
    \frac{1}{2}
    \exp\left\{\left(\frac{1}{K_p} -\frac{1}{K}\right)G_{0r}^{(-)} 
	 -\sigma\right\}
    -
    \frac{1}{2}
    \exp\left\{\left(\frac{1}{K_p} -\frac{1}{K}\right)G_{0r}^{(+)} 
	 -\sigma\right\},
      \end{eqnarray}
  where $G_{0r}$ and $G_{0r}^{(\pm)}$ are the lattice Green
  functions,
  \begin{eqnarray}
   G_{0r} &=&
    \frac{1}{N}\sum_{\Vect{k}}
    \frac
    {1 -\cos \Vect{k}\cdot\Vect{r}}
    {4 -2\cos k_x -2\cos k_y}\\
    G_{0r}^{(\pm)} &=& 
     \frac{1}{N} 
      \sum_{\Vect{k}}\frac{(1-\cos k_1)
      (1 \pm \cos \Vect{k}\cdot\Vect{r})}{2 -\cos k_1 -\cos k_2}.
  \end{eqnarray}
  Now, applying the special condition $K=K_p$, we can obtain these
  correlation functions as follows,
  \begin{eqnarray}
   C_{\rm p} = \e^{-K_p^{-1} r/2},\,\,\,
   C_{\rm le} = \e^{K_p^{-1}},\,\,\,
   C_{\rm ch} = 0.\label{eqn:spinsol}
  \end{eqnarray}
  Surprisingly, the chirality correlation function vanishes under the
  condition $K=K_p$.
  The chirality degrees of freedom behave perfectly independently from
  place to place.

 \subsection{Gauge transformation}
 From comparison of the results of the spin wave approximation with the
 exact results, we can investigate the difference between the model with
 periodicity and the model without periodicity.
 In this section, we calculate the exact solutions of $C_{\rm p}$,
 $C_{\rm le}$ and $C_{\rm ch}$ for models with periodicity using gauge
 transformation under the special condition $K=K_p$.
 After the derivations of the exact solutions of correlation functions,
 we study the asymptotic forms of the exact solutions at low
 temperatures and compare them with the solutions in the previous
 section using the spin wave approximation.\\
 The correlation functions which we introduced in Section 3.1 are
 invariant under the gauge transformation, and therefore the same
 calculation method as for the internal energy (\ref{eqn:exact1}) in
 Section 2 can be applied to the calculation of the correlation
 functions.
 As a consequence, we obtain the following expression under the special
 condition $K=K_p$,
 \begin{eqnarray}
  C_{\rm p} =
   \left(\frac{I_1(K)}{I_0(K)}\right)^r,\,\,\,
  C_{\rm le} &=& \left(\frac{I_1(K)}{I_0(K)}\right)^2,\,\,\,
  C_{\rm ch} = 0. \label{eqn:genmitu}
 \end{eqnarray}
 When $K=K_p$, the chirality correlation vanishes
 exactly and this solution is consistent with the result of spin wave
 approximation (\ref{eqn:spinsol}).
 At low temperature, the asymptotic behaviors of $C_{\rm p}$ and $C_{\rm
 le}$ in eqn (\ref{eqn:genmitu}) are written as follows,
 \begin{eqnarray}
  C_{\rm p} \sim {\rm e}^{-K_p^{-1}r/2},\,\,\,
  C_{\rm le}\sim {\rm e}^{-K_p^{-1}}.\label{eqn:exactc2}
 \end{eqnarray}
 Equation (\ref{eqn:exactc2}) is also consistent with the solution of
 spin wave approximation expressed in eqn (\ref{eqn:spinsol}). 
 This comparison suggests that under the special condition $K=K_p$, the
 system forms a characteristic structure which is not influenced by the
 vortices.
 The analysis of the Villain model in the following reinforces this
 picture.

 \subsection{Villain model}
 The Hamiltonian of the spin wave approximation (\ref{eqn;sw}) neglects
 periodicity included in the cosine term (\ref{eqn:HGG}).
 If we need to describe the system at low temperatures more accurately,
 it should be necessary to consider periodicity in the Hamiltonian.
 Accordingly, we consider the Villain model which adds the periodicity
 to the spin wave Hamiltonian
 \begin{eqnarray}
  Z_V = \int_0^{2\pi}\prod_i\frac{\dei\theta_i}{2\pi}
   \sum_{\{m_{ij}=-\infty\}}^{\infty}
   \e^{-\frac{K}{2}\sum_{\left\langle i,j\right\rangle}
   (\theta_i -\theta_j -A_{ij} -2m_{ij}\pi)^2}.\label{eqn:Villaindis}
 \end{eqnarray} 
 where $m_{ij}$ is an integer between $-\infty$ and $\infty$.\\
 The random phase shift (\ref{eqn:bond:gauss}) also neglects the
 periodicity of modulo $2\pi$.
 Therefore we also add $2m_{ij}\pi$ into the probability distribution of
 $A_{ij}$ (\ref{eqn:bond:gauss}).
 The configurational average is described as
 \begin{eqnarray}
  [\cdots] = \int_0^{2\pi}\prod_{\left\langle i,j\right\rangle}
   \frac{\dei A_{ij}}{\sqrt{2\pi K_p^{-1}}}(\cdots)
   \sum_{\{m_{ij}=-\infty\}}^{\infty}
   \e^{-\frac{K_p}{2}(A_{ij}-2m_{ij}\pi)^2}.\label{eqn:Villaincon}
 \end{eqnarray}
 From eqns (\ref{eqn:Villaindis}) and (\ref{eqn:Villaincon}) and the
 gauge transformation, we can calculate the correlation functions
 discussed in the previous section.
 We first consider the gauge-invariant correlation function $C_{\rm
 p}$.
 Applying the gauge transformation and the condition $K=K_p$,
 we obtain the following form,
 \begin{eqnarray}
  C_{\rm p} &=&
   \left[\left\langle 
       \exp{\im(\theta_0 -\theta_r 
       -\sum_{k=1}^{r} A_{k-1\,k})} \right\rangle
   \right] \nonumber\\
  &=&
   \int_0^{2\pi}\prod_{\left\langle i,j\right\rangle}
   \frac{\dei A_{ij}}{\sqrt{2\pi K_p^{-1}}}
   \sum_{\{m_{ij}=-\infty\}}^{\infty}\nonumber\\
  &&\times
   \int_0^{2\pi}\prod_i\frac{\dei\theta_i}{2\pi}
   \e^{\im(\theta_0 -\theta_r -\sum_{k}A_{k-1\,k})}
   \e^{-\frac{K_p}{2}\sum_{\left\langle i,j \right\rangle}
   (\theta_i -\theta_j -A_{ij} -2m_{ij}\pi)^2}.\label{eqn:villanpart}
 \end{eqnarray}
 To integrate over $A_{ij}$, we use the following relation,
 \begin{eqnarray}
  \int_{0}^{2\pi}
   \frac{\dei A_{ij}}{\sqrt{2\pi K_p^{-1}}}
   \sum_{m_{ij} =-\infty}^{\infty}
   \e^{-\frac{K_p}{2}   
   (\theta_i -\theta_j -A_{ij} -2m_{ij}\pi)^2}&& \nonumber\\
   &&\hspace*{-4cm}=
   \int_{-\infty}^{\infty}
   \frac{\dei A_{ij}}{\sqrt{2\pi K_p^{-1}}}
   \e^{-\frac{K_p}{2}   
   (\theta_i -\theta_j -A_{ij})^2}\label{eqn:gaussform}
 \end{eqnarray}
 Substituting eqn (\ref{eqn:gaussform}) into eqn (\ref{eqn:villanpart}),
 we obtain the gauge-invariant correlation function as
 \begin{eqnarray}
  C_{\rm p} = \e^{-K_p^{-1} r/2},
 \end{eqnarray}
 where we have used the relation $K = K_p$.
 The result of this calculation agrees with the one of the spin
 wave approximation (\ref{eqn:spinsol}) and the asymptotic form of the
 exact solution (\ref{eqn:exactc2}) under the condition $K=K_p$.
 The local energy correlation function and the chirality correlation
 function are derived by the same calculation as in the gauge invariant
 correlation.
 These results are written by simple formulas as 
 \begin{eqnarray}
  C_{\rm le} = \e^{-K_p^{-1}},\,\,\,
  C_{\rm ch} = 0.
 \end{eqnarray}
 Those two correlation functions also agree with the results of the spin
 wave approximation (\ref{eqn:spinsol}) and asymptotic exact solution
 (\ref{eqn:exactc2}) under the condition $K = K_p$.
 We may conclude from these results as follows.
 The spin wave approximation neglects periodicity, and the results of
 the spin wave approximation do not include the effects of vortices.
 The results for the Villain model in contrast include the effects of
 vortices.
 Accordingly, the agreement of the results of spin wave approximation
 with those of the Villain model suggests the absence of the effects of
 vortices on the correlation functions we calculated under the condition
 $K=K_p$.
 Vortices are perfectly irrelevant to the gauge invariant correlation
 functions.

 \section{Distribution functions}
 Irrelevance of vortices in the gauge invariant correlation functions on
 the Nishimori line can be verified by using a distribution function
 with gauge invariance.
 This gauge invariant distribution function measures the distribution of
 the random phase twist between site $i$ and $j$.
 The definition is
 \begin{eqnarray}
  P(x) = \bigg[\big\langle\delta
   \left(x-(\theta_i -\theta_j -A_{ij})\right)
   \big\rangle\bigg].\label{eqn:px}
 \end{eqnarray}
 Using the spin wave approximation and the Villain model, we obtain the
 comparable results as in the gauge invariant correlation functions when
 $K=K_p$.
 The results using the spin wave approximation and the Villain model
 are, respectively,
 \begin{eqnarray}
  P_{\rm sw}(x) &=& \sqrt{\frac{K_p}{2\pi}}
   \exp\left(-\frac{K_p}{2}x^2\right)\label{eqn:psw}\\
  P_{\rm Villain}(x) &=& \sqrt{\frac{K_p}{2\pi}}
   \exp\left(-\frac{K_p}{2}x^2\right),\label{eqn:pvil}
 \end{eqnarray}
 the latter being the exact solution.
 These two expressions are completely identical.\\
 The distribution function of two bonds can also be calculated by
 the spin wave approximation and the Villain model.
 We first show the definition of the two-bond distribution function
 \begin{eqnarray}
  P(x,y) = \bigg[\big\langle
   \delta\left(x-(\theta_k -\theta_l -A_{kl})\right)\cdot
   \delta\left(y-(\theta_m -\theta_n -A_{mn})\right)
   \big\rangle\bigg].\label{eqn:two}
 \end{eqnarray}
 Equation (\ref{eqn:two}) is calculated by using the spin wave
 approximation and the Villain model for $K=K_p$ as,
 again the latter being the exact solution,
 \begin{eqnarray}
  P_{\rm sw}(x,y) &=& \frac{K_p}{2\pi}
   \exp\left(-\frac{K_p}{2}(x^2 +y^2)\right)\label{eqn:p2sw}\\
  P_{\rm Villain}(x,y) &=& \frac{K_p}{2\pi}
   \exp\left(-\frac{K_p}{2}(x^2 +y^2)\right).\label{eqn:p2vil}
 \end{eqnarray}
 Similarly to eqns (\ref{eqn:psw}) and (\ref{eqn:pvil}), these results
 agree with each other.
 We therefore see immediately that vortices have no effects on the
 distribution function (\ref{eqn:two}).
 It also follows that two-bond variables behave independently because
 eqns (\ref{eqn:p2sw})/(\ref{eqn:p2vil}) are just products of eqns
 (\ref{eqn:psw})/(\ref{eqn:pvil}).\\
 It is instructive to consider another two-bond extension,
 \begin{eqnarray}
  P'(x,y) = \bigg[\big\langle
   \delta\left(x-(\theta_k -\theta_l -A_{kl})\right)
   \big\rangle\cdot
   \big\langle
   \delta\left(y-(\theta_m -\theta_n -A_{mn})\right)
   \big\rangle\bigg].\label{eqn:pdash}
 \end{eqnarray}
 Although eqn (\ref{eqn:pdash}) is gauge invariant, we can not calculate
 the exact solution even if we apply gauge transformation to eqn
 (\ref{eqn:pdash}).
 Nevertheless, we can obtain the result by the spin wave approximation
 as 
 \begin{eqnarray}
  P'_{\rm sw}(x,y) = 
   \frac{K_p}{2\pi \sqrt{1 -A^2}}
   \exp\left(-\frac{K_p}{2(1-A^2)}(x^2 +y^2 -2Axy)\right),\label{eqn:A}
 \end{eqnarray}
 where $A\equiv G_{kn} -G_{lm} -G_{km} +G_{ln}$.
 In contrast to eqn (\ref{eqn:two}), the distribution function
 (\ref{eqn:pdash}) can not be written as the products of eqn
 (\ref{eqn:psw}) under the condition $K=K_p$.
 Bond variables are correlated in this sense.
 Inequivalence of eqns (\ref{eqn:two}) and (\ref{eqn:pdash}) is a highly
 non-trivial fact.
 If we measure the two-bond correlation within a given system as in eqn
 (\ref{eqn:two}), the result becomes uncorrelated: $P(x,y)=P(x)P(y)$.
 If we, on the other hand, look at the correlation of the thermal
 averages of single-bond distribution function at two different
 locations as in eqn (\ref{eqn:pdash}), these are correlated:
 $P'(x,y)\neq P(x)P(y)$.
 Physical significance of this counter-intuitive results needs further
 clarification.

 \section{Summary and discussions}
 In this paper, we have investigated the properties of the gauge glass
 on the Nishimori line through the gauge invariant quantities.
 Firstly we calculated the gauge invariant correlation functions using
 the spin wave approximation, gauge transformation and the Villain
 model.
 From these calculations, we found that the results of the spin wave
 approximation are identical to the results by the method taking
 periodicity into account.
 Thus, we expect that the periodicity or vortices has no influence on
 the system (gauge invariant quantities to be precise) under the
 condition $K=K_p$. 
 The effect of vortices can be neglected on the Nishimori line.\\
 Next, we calculated the gauge invariant distribution functions using
 the spin wave approximation and the Villain model.
 We found the same results as for the gauge invariant correlation
 functions.
 Moreover, we showed that the distribution function (\ref{eqn:two}) is
 independently distributed at each bond. 
 An variant of eqn (\ref{eqn:two}) was also calculated by using the
 spin wave approximation.\\
 We may conclude from these results as follows.
 In the two-dimensional gauge glass, the effect of vortices plays an
 important role because the phase transition of the model is described
 by vortex-pair unbinding.
 Accordingly, the spin wave approximation which neglects the effect of
 vortices is considered in general not to be able to  describe the
 singularity related to the phase transition.
 However, our analysis based on the gauge invariant quantities predicts
 that under the condition $K=K_p$, the singularity related to the phase
 transition vanishes for gauge invariant correlation functions even if
 we consider the effect of vortices in the system.
 Thus, the renormalization group arguments \cite{CD00}\cite{T96}, which
 analyze the effects of vortices asymptotically, should be reconsidered
 seriously under the present perspective.
 Further investigation of the property of the Nishimori line is
 necessary.

\end{document}